\documentclass[acmsmall]{acmart}

\usepackage{amsmath}

\setcopyright{none}
\copyrightyear{}
\acmYear{}
\acmDOI{}

\acmConference[]{}{}{}
\acmBooktitle{}
\acmPrice{}
\acmISBN{}

\begin{document}

\title[1000 Pupil Segmentations in a Second]{1000 Pupil Segmentations in a Second using Haar Like Features and Statistical Learning}

\author{Wolfgang Fuhl}
\email{wolfgang.fuhl@uni-tuebingen.de}
\orcid{0000-0001-7128-298X}
\affiliation{%
	\institution{University T\"ubingen}
	\streetaddress{Sand 14}
	\city{T\"ubingen}
	\state{Baden W\"urttemberg}
	\country{Germany}
	\postcode{72076}
}

\renewcommand{\shortauthors}{Fuhl}

\begin{abstract}
  In this paper we present a new approach for pupil segmentation. It can be computed and trained very efficiently, making it ideal for online use for high speed eye trackers as well as for energy saving pupil detection in mobile eye tracking. The approach is inspired by the BORE and CBF algorithms and generalizes the binary comparison by Haar features. Since these features are intrinsically very susceptible to noise and fluctuating light conditions, we combine them with conditional pupil shape probabilities. In addition, we also rank each feature according to its importance in determining the pupil shape. Another advantage of our method is the use of statistical learning, which is very efficient and can even be used online.  \url{https://atreus.informatik.uni-tuebingen.de/seafile/d/8e2ab8c3fdd444e1a135/?p=\%2FStatsPupil\&mode=list}.
\end{abstract}




\begin{teaserfigure}
  \includegraphics[width=\textwidth]{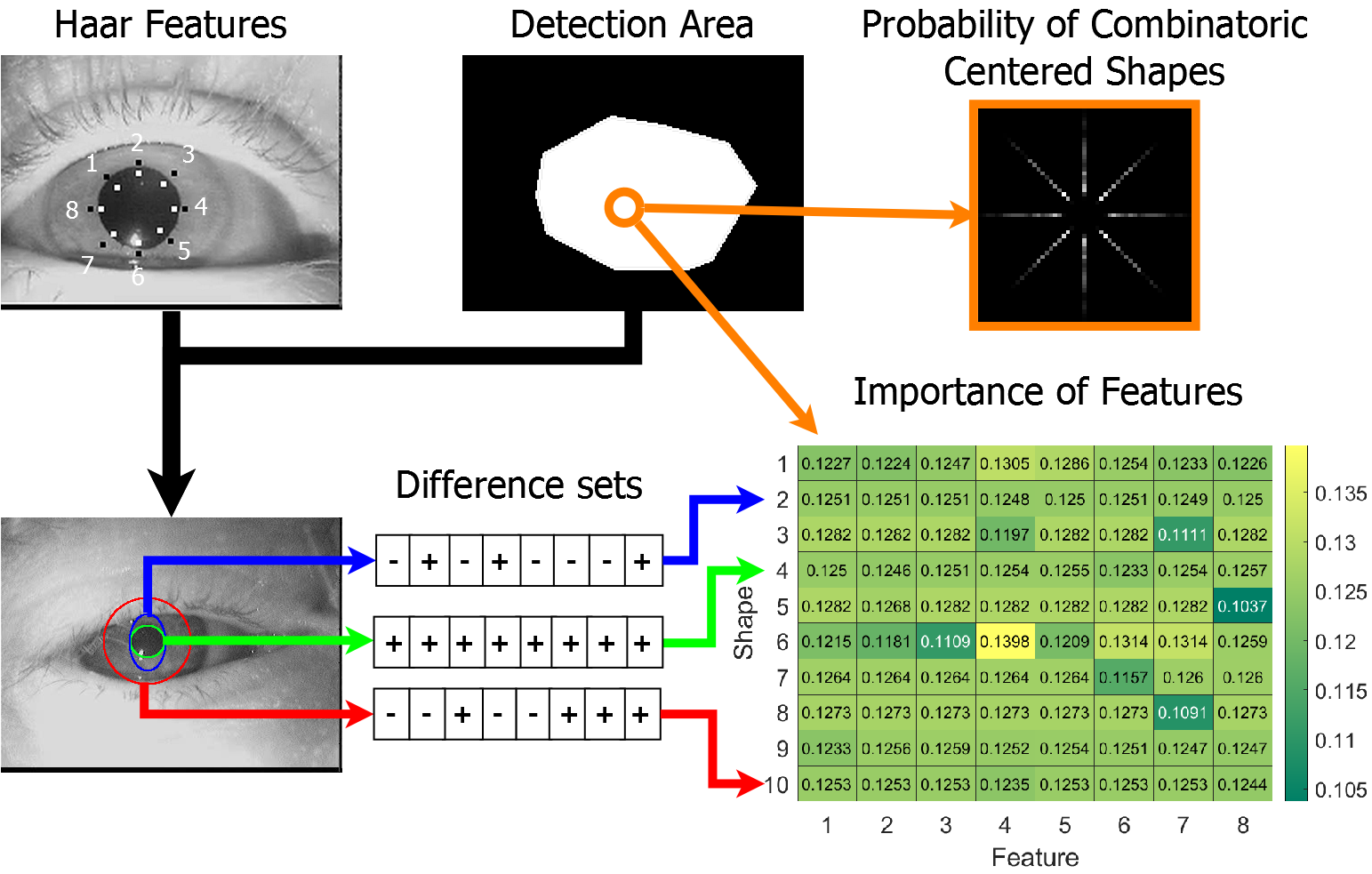}
  \caption{We introduce tiny Haar features which follow elliptical shapes for pupil segmentation. Together with a detection area, a shape conditioned probability distribution as well as statistical feature waiting.}
  \label{fig:teaser}
\end{teaserfigure}

\maketitle

\section{Introduction}
The plethora of image based eye tracking~\cite{duchowski2002breadth,NNETRA2020} applications has continued to rise in recent years. The most important areas of application are currently driver monitoring~\cite{brandt2004affordable}, virtual reality~\cite{meissner2019combining}, augmented reality~\cite{naspetti2016automatic}, medicine~\cite{032017,0320170,Bahmani2016}, market research~\cite{meissner2019combining,piumsomboon2017exploring}, remote support~\cite{silva2019eye}, human computer interaction~\cite{UMUAI2020FUHL,majaranta2014eye}, supportive explanation models for computer vision models~\cite{yun2013studying}, and many more. Of course, not only pupil detection is important for this, but also scan path analysis~\cite{C2019,FFAO2019}, eye movement classification~\cite{FCDGR2020FUHLARX,FCDGR2020FUHL,fuhl2018simarxiv,ICMIW2019FuhlW1,ICMIW2019FuhlW2,EPIC2018FuhlW,MEMD2021FUHL}, visualizations~\cite{ROIGA2018,ASAOIB2015,TCKWGJRWE2015,AGAS2018} and validations of the approaches and models used~\cite{ICMV2019FuhlW,NNVALID2020FUHL}.

Those diverse application areas bring different image based challenges~\cite{WTTE032016,WTCKWE092015,062016,WTCDAHKSE122016,WTCDOWE052017,WF042019} and challenging resource restrictions~\cite{VECETRA2020,WDTTWE062018,ETRA2018FuhlW}. Some of these image based challenges are changing illumination conditions, reflections on glasses, make up, recording errors, and high off axial pupil positions. In addition, the diversity of people using eye tracking devices also rise new challenges like deformed pupils~\cite{WTTE032016,WTCKWE092015,CORR2017FuhlW1,CORR2017FuhlW2,CORR2016FuhlW} which occurs after eye surgery~\cite{dhir2010effect}.

Other challenges in eye tracking are different recording techniques like RGB and NIR imaging. While NIR is mostly used in head mounted~\cite{062016} eye trackers, RGB imaging is still used in remote~\cite{WDTE092016} eye tracking and especially web cam based eye tracking~\cite{papoutsaki2017searchgazer}. Due to the current situation with the Covid pandemic, web cam based eye tracking becomes more and more important for market research~\cite{papoutsaki2017searchgazer,meissner2019combining} and scientific studies~\cite{papoutsaki2017searchgazer}.

Nowadays, the gaze signal alone is also no longer sufficient, as the eye provides a variety of other sources of information. These are pupil response to cognitive load~\cite{chen2014using}, pupil shape for eyeball regression~\cite{NNETRA2020}, and eyelids to determine a person's fatigue~\cite{WTDTWE092016,WTDTE022017,WTE032017}. The cognitive load is very interesting for the detection of mental disorders~\cite{najmi2015effects} or ranking a persons performance capability~\cite{maksimenko2018increasing}. Eye ball regression is used to improve the robustness of eye trackers against drifts of the device and to improve the accuracy~\cite{sugano2015self}. The fatigue detection of a person is important for critical applications like driving~\cite{xu2018real}, flying~\cite{naeeri2019analyzing}, flight surveillance~\cite{mckinley2011evaluation}, and many more.

Due to the progress in eye tracker technology so far, mobile applications~\cite{krafka2016eye}, long-term studies~\cite{vidal2012wearable}, high speed eye tracking for fundamental research~\cite{hosp2020remoteeye}, and the consumer market such as computer games~\cite{jonsson2005if} as well as the privacy aspects of eye tracking~\cite{RLDIFFPRIV2020FUHL,GMS2021FUHL} are becoming more and more important. For this it is necessary that the algorithms can be used as resource saving and robust as possible~\cite{WDTTWE062018,ETRA2018FuhlW,VECETRA2020,AAAIFuhlW,NNPOOL2020FUHL} to consume as little energy as possible in mobile applications~\cite{krafka2016eye}, to guarantee the real-time capability in high speed eye tracking~\cite{hosp2020remoteeye}, and not to waste computing capacity which is needed for computer games~\cite{meng2018kernel}.

In this paper we present a resource-sparing approach, which is inspired by CBF~\cite{WDTTWE062018} and BORE~\cite{ETRA2018FuhlW}. Our approach was developed with the main features of cheap execution and easy training. The features used are Haar features~\cite{viola2001rapid} which can be computed very efficiently. In addition, we increase the computation of Haar features via down scaling the images instead of computing the integral image. Another important feature of our algorithm is the use of conditional pupil ellipse distributions, which allow to consider only the ellipses that can occur. As a training method we use statistical learning, which has a complexity of $O(n)$ and can be computed very efficiently. In this way, our detector can even be personalized and used optimally for individuals using the minimum resources.

Contribution of this work to the state of the art:
\begin{enumerate}
	\item We define the features in comparison to BORE~\cite{ETRA2018FuhlW} and CBF~\cite{WDTTWE062018} generalized as Haar features. In CBF~\cite{WDTTWE062018} and BORE~\cite{ETRA2018FuhlW} only direct pixel comparisons were used, we use the difference of areas. 
	\item Our approach is the first to use ellipse parameter conditional probability distributions for ellipse selection. This avoids unnecessary checks of ellipse points, as is the case in CBF~\cite{WDTTWE062018} and BORE~\cite{ETRA2018FuhlW}.
	\item Compared to BORE~\cite{ETRA2018FuhlW} and CBF~\cite{WDTTWE062018} we use index tables whereby each feature has to be evaluated only once. This further reduces resource consumption and, in combination with the precomputed indexes already presented in CBF~\cite{WDTTWE062018} and BORE~\cite{ETRA2018FuhlW}, further speeds up the process.
	\item Our approach is simply trained on the occurrence statistics. This procedure is much more resource efficient than the unsupervised learning and evaluation of all possible combinations as done in BORE~\cite{ETRA2018FuhlW}. It is also much faster than the random combination evaluation used in CBF~\cite{WDTTWE062018}.
	\item Using feature weighting, our approach also has the ability to find ellipses that are not fully present in the image. 
	\item Compared to BORE~\cite{ETRA2018FuhlW} and CBF~\cite{WDTTWE062018}, our approach segments the pupil explicitly. For BORE~\cite{ETRA2018FuhlW}, there is only one experimental implementation for ellipse extraction and in the case of CBF~\cite{WDTTWE062018}, only the pupil center.
	\item Compared to the Tiny CNNs~\cite{VECETRA2020,ICMV2019FuhlW}, our approach has significantly reduced runtime and hence resource consumption. In addition, we only need a fraction of time to train our model as well as no GPU to execute the teacher network.
\end{enumerate}

\section{Related work}
Since pupil tense has evolved in different directions, we divide the related work into three areas. These are classical computer vision approaches, deep neural networks, and resource saving machine learning approaches.

\subsection{Classical Computer Vision Approaches}
In the field of pupil detection and exact pupil center determination, the first major breakthrough came with the use of cate images~\cite{swirski2012robust}. Previously, adaptive thresholds were used~\cite{haro2000detecting}. A major disadvantage of edge images are their susceptibility to noise and motion blur. Therefore, edge filtering methods were introduced~\cite{WTTE032016,WTCKWE092015,WTCDAHKSE122016}, which suppress noise and pass only relevant edge segments. In addition to this, angular integral projection function~\cite{WTCKWE092015} and also blob detection~\cite{WTTE032016} were used. Another improvement in pupil shape reconstruction was the evaluation of individual segments~\cite{javadi2015set}. Alternative to edge detection, the radial symmetry transform was used to detect the pupil center~\cite{martinikorena2018fast}.

\subsection{Deep Neural Networks}
With the advent of convolutions in neural network and the success in the field of image processing, these CNNs were also used for pupil detection and segmentation as well as they are continuously refined~\cite{NORM2020FUHL,RINGRAD2020FUHL}. The first window-based approach was PupilNet~\cite{CORR2016FuhlW,CORR2017FuhlW2}, running in real time on a single CPU core only. Later, large residual networks were also used~\cite{CAIP2019FuhlW,ICCVW2018FuhlW} and puplished together with huge annotated data sets as well as generative adversarial networks~\cite{ICCVW2019FuhlW} were used. The first U-Net with interconnections was poposed with DeepVOG~\cite{yiu2019deepvog}. New loss formulations regarding the pupil shape where proposed in \cite{chaudhary2019ritnet}. Additional to those loss formulations an L1 loss connected to the central part of a fully connected convolutional network was proposed in \cite{kothari2020ellseg}.

\subsection{Resource Saving Machine Learning}
The first real-time machine learning methods combined with simple features were introduced by PupilNet~\cite{CORR2016FuhlW,CORR2017FuhlW2} and continued BORE~\cite{ETRA2018FuhlW} and CBF~\cite{WDTTWE062018}. BORE~\cite{ETRA2018FuhlW} is capable of non-supervised learning and self-optimization. CBF~\cite{WDTTWE062018}, on the other hand, uses random ferns and pixel comparisons to determine the center of the pupil. Also the supervised decent method (SDM) was used for the regression of the pupil center in remote images~\cite{larumbe2018supervised}. However, this has the disadvantage of being highly dependent on the mean shape, which we will show in our later evaluation in combination with landmarks for segmentation on head mounted images. Another approach was created using the teacher, student training method~\cite{VECETRA2020,ICMV2019FuhlW}. These tiny cnns~\cite{VECETRA2020,ICMV2019FuhlW} are very robust and were successfully used across datasets. The paper reported a runtime of 16ms but the provided nets only have a runtime of 4-8ms on a CPU core. In addition, they learn to evaluate the accuracy and give therefore a validity of the pupil ellipse~\cite{ICMV2019FuhlW}.

The approach presented by us is also to be classified in this category. This is due on the one hand to the fact that our approach uses statistical learning and on the other hand to the resource-saving use of our method. In addition to these properties, our approach can also be trained very fast and resource-saving.

\section{Method}
Our approach uses statistical learning and by this it is enough to look at each training sample three times to create a detector. The first step for training our detector is to create the search area ($x,y$ tuples) and the ellipses expected there ($el_i$). This gives us the set $EL$ which stores all ellipses $el_i$ for each position $x,y$. Reformulated as a conditional probability distribution $EL$ corresponds to the probability $P$ of ellipse $el_i$ under the condition to be at position $x,y$ and thus Equation~\ref{eq:area}.

\begin{equation}
EL=P(el_i|x,y)
\label{eq:area}
\end{equation}

To calculate EL, we need one pass of the training data. In the second step, we reduce EL to speed up our detector and reduce over fitting. For this, we represent each ellipse as eight landmarks (See Figure~\ref{fig:teaser}) and round them to integers. For reduction, all ellipses with the same landmark distances are combined into one ellipse with a maximum deviation of one pixel per landmark. This gives us the reduced conditional probability distribution $\widehat{EL}$.

The next step is to create our feature extractors from the landmarks. For this we use Haar features. Instead of computing the area differences in the integral images, we use the difference of pixels in downscaled images. In the second pass of the training set, for each ellipse in $\widehat{EL}$ we store all occurrences of the eight differences $d_j$. Since the set of eight differences to each ellipse is very large, we want to reduce it. To reduce this set, we compute the best five difference sets $d_j$ for each positive probability in $\widehat{EL}$, noting here that one $d_j$ are eight differences one for each landmark. For this we use the mean shift clustering with a maximum of five clusters. Reformulated as a conditional probability distribution, we thus have positive probabilities for five difference sets $d_j$ under the condition of the probability of an ellipse at some position ($\widehat{EL}=P(\widehat{el}_i|x,y)$) and thus Equation~\ref{eq:diff}.

\begin{equation}
D=(d_j|P(\widehat{el}_i|x,y))
\label{eq:diff}
\end{equation}

In the last run, the individual differences or landmarks are now weighted with respect to their robustness. These feature weights $f_j$ are computed in the third run of the training set. For this we use the difference set $d_j$ with the minimum distance of the landmark differences and weight $f_j$ positively if the sign matches and negatively if the sign differs. Based on this we statistically weight the reliability of each feature. After the pass each feature $f_j$ is normalized to sum to one and also form a probability distribution. This gives us a similar conditional probability distribution as for the difference sets and thus Equation~\ref{eq:feature}.

\begin{equation}
F=(f_j|P(\widehat{el}_i|x,y))
\label{eq:feature}
\end{equation}

To use the detector, the landmark differences to the ellipses must be calculated at all possible positions. Then the minimum difference $j$ to the difference sets $d_j$ is calculated and the deviation is weighted by the feature weights $f_j$. The final ellipse and position is then the global minimum and described in Equation~\ref{eq:detector}.

\begin{equation}
\underset{P(\widehat{el}_i|x,y))}{argmin}(\underset{j}{argmin}\sum_{LM=1}^{8} abs(d_j(LM)-\tilde{d}(LM))*f_j(LM))
\label{eq:detector}
\end{equation}

In Equation~\ref{eq:detector} $LM$ are the haar features, $d_j$ is the difference set, $\tilde{d}$ is the set of differences in the input image for ellipse $\widehat{el}_i$ at position $x,y$ and $f_j$ is the corresponding feature weight. Overall Equation~\ref{eq:detector} searchs for the minimum difference of the eight landmark positions in the entire input image. If there are multiple equally good positions we use the conditional probability distribution $\widehat{EL}$ to select the most probable ellipse and position. While this already provides a very efficient detector, further optimizations are necessary like the precalculation of all indexes in the image, as it was already presented with BORE~\cite{ETRA2018FuhlW} and CBF~\cite{WDTTWE062018}. Also, all differences are indexed to calculate differences at each position only once.

\section{Evaluation}
The data used for the evaluation are the segmented pupils of \cite{CAIP2019FuhlW}. The data set consists of two files \"p1\_image.mp4\" and \"p2\_image.mp4\" with an image resolution of $192\times144$. The first file contains data taken in a driving simulator and the second file contains images from real world driving. Because of this there are no reflections or strong light fluctuations in the data of the first file, so it contains much simpler images. Therefore, we decided to use the first file with more than 500,000 frames as the training data set. The second file with more than 350.000 frames and the much more challenging images is used as the evaluation data set.

For the data augmentation, we used up to 20\% random noise, as well as reflections with an intensity up to 20\%, where the reflections are calculated from randomly selected images. Also, we randomly changed the contrast of the image in the range of -40 to 40. In addition, we shifted the image randomly in a range of -10 to 10 pixels as well as we used zooming with a random factor in the range of 0.8 to 1.2. For the TinyCNNs~\cite{VECETRA2020}, this was done online during training. For all other approaches, the data augmentation was computed in advance resulting in five images from each frame. Of course, the image could also occur without augmentation.

Since we trained our approach once with the real data and once with the simulator data, we also give here the details for our approach. First, we used the simulator of \cite{NNETRA2020} and inverted the images so that the pupil is dark. Then we selected the data based on the pupil ellipse, which matched the pupil ellipses in the normal training set. For the data augmentation, we used the same approaches as for the training on the real data except for adjusting the contrast of the background and the pupil of the simulated images. Here we used the differences from the training set first to adjust the contrast.

The hardware used for training and running the final models consists of an Intel i5-4570 CPU running at 3.2 GHz. The system has 16GB of DDR4 memory and an NVIDIA 1050 TI with 4GB of GDDR5 memory. The GPU was only used for training the TinyCNNs~\cite{VECETRA2020}. All runtime analyses were performed on one CPU core.

For a comparison with the state of the art, we use ElSe as a representative of edge-based approaches, BORE as a resource saving alternative, the TinyCNNs pre-trained on LPW~\cite{tonsen2016labelled} and provided by the authors as well as two newly trained TinyCNNs on the presented training data, and SDM~\cite{xiong2013supervised} for landmark detection also as a resource sparing alternative.

\begin{figure}
	\centering
	\includegraphics[width=0.49\textwidth]{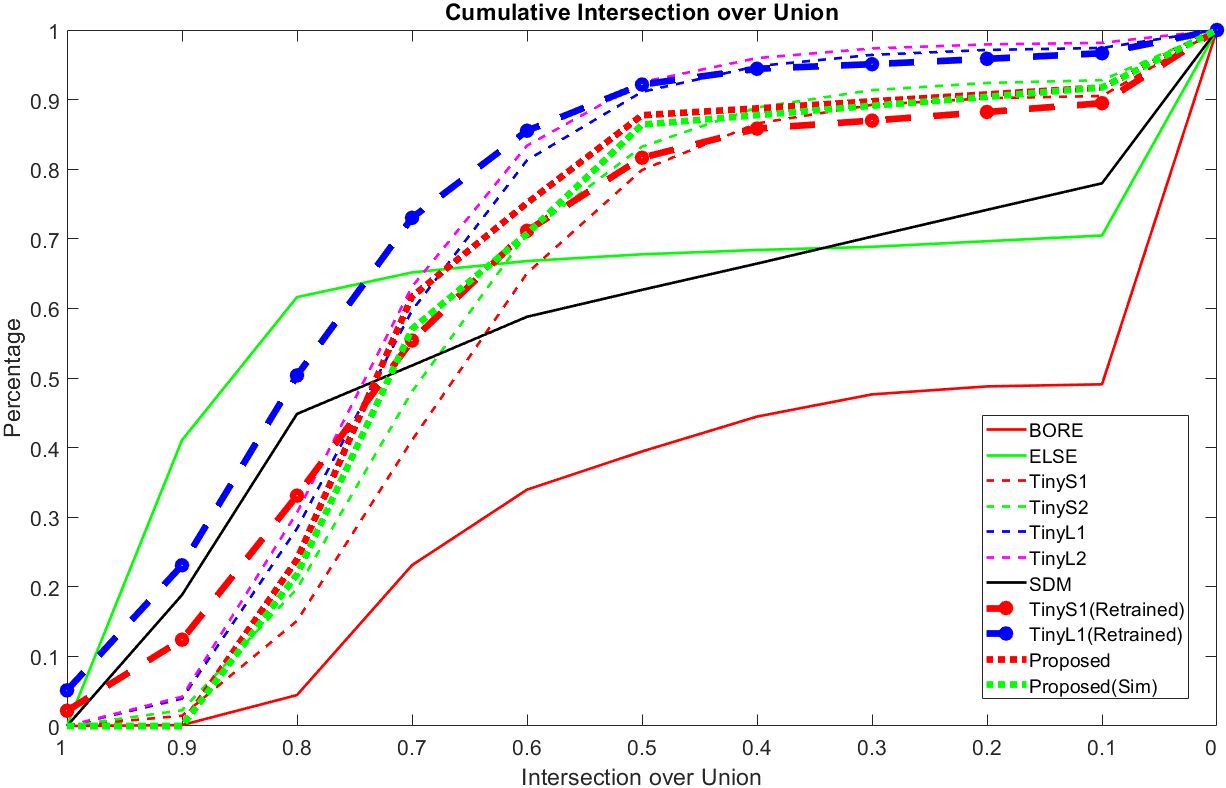}
	\includegraphics[width=0.49\textwidth]{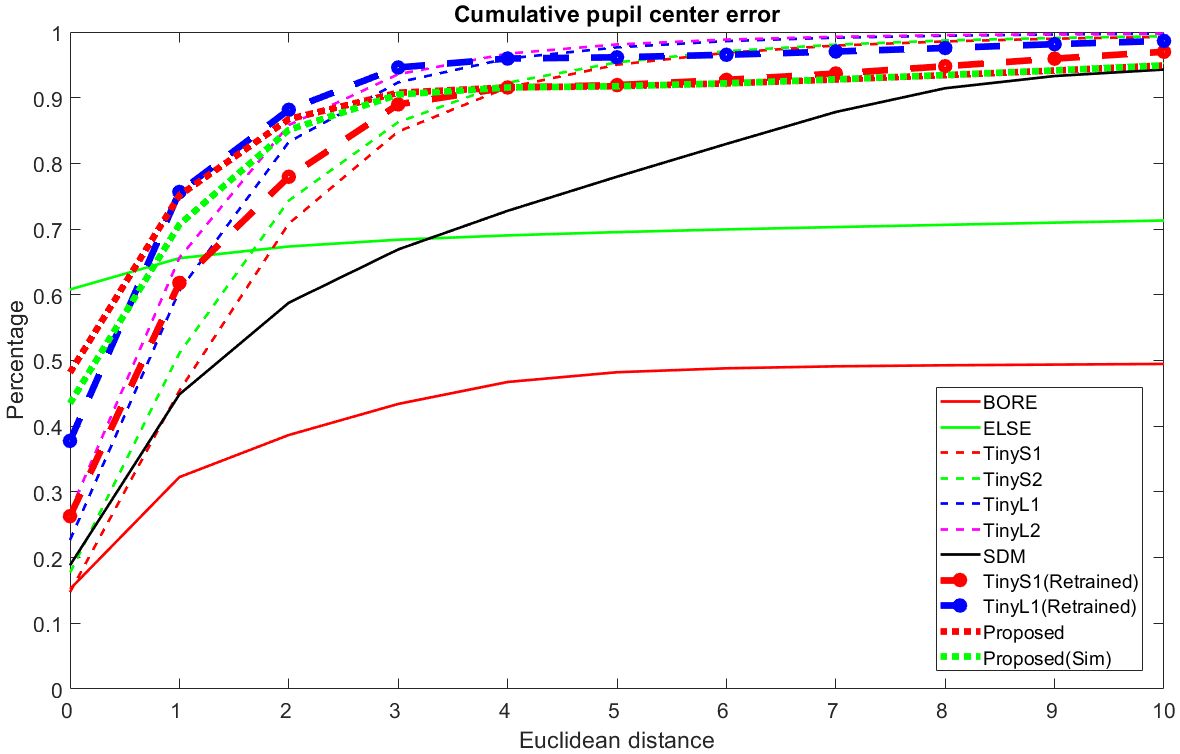}
	\caption{The cumulative intersection over union results for multiple approaches on the left. On the right is the cumulative euclidean pupil distance error for multiple approaches.}
	\label{fig:cumuplot}
\end{figure}

\begin{figure}
	\centering
	\includegraphics[width=0.24\textwidth]{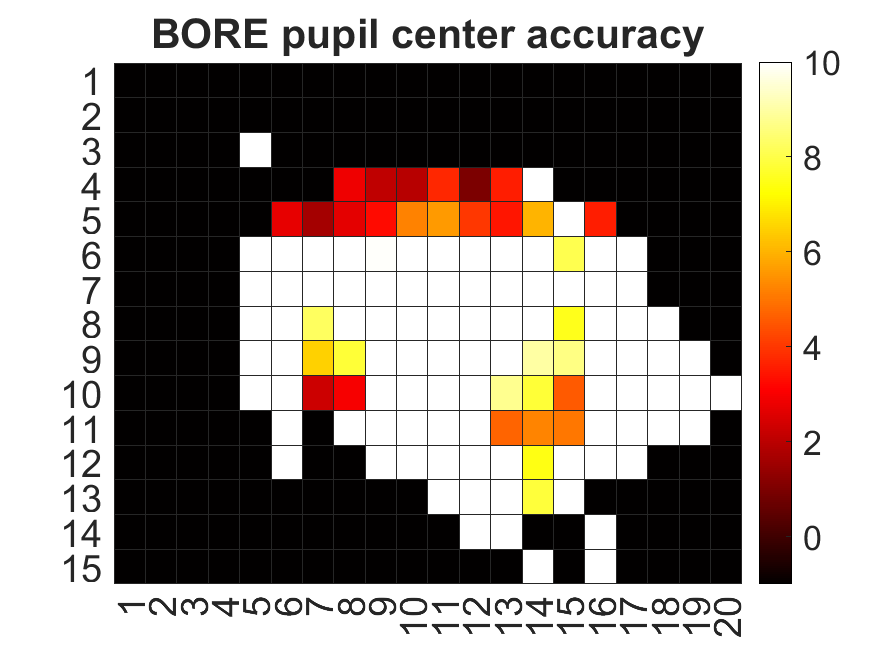}
	\includegraphics[width=0.24\textwidth]{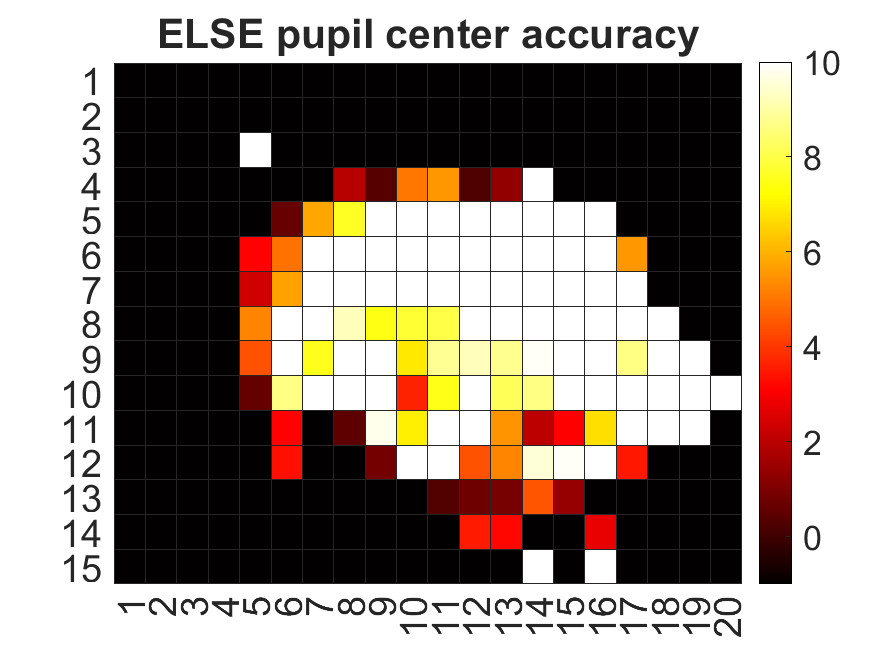}
	\includegraphics[width=0.24\textwidth]{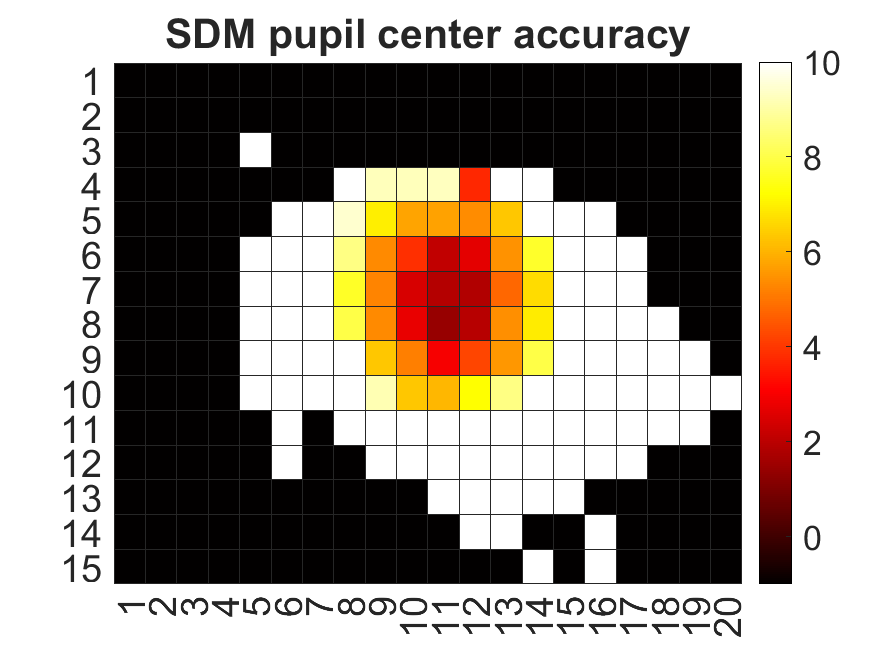}
	\includegraphics[width=0.24\textwidth]{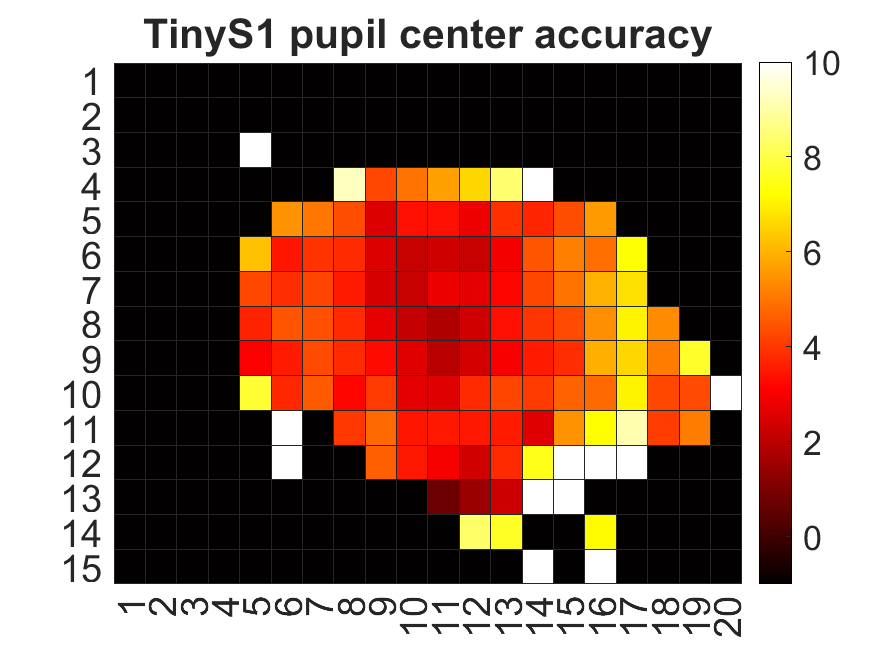}
	\includegraphics[width=0.24\textwidth]{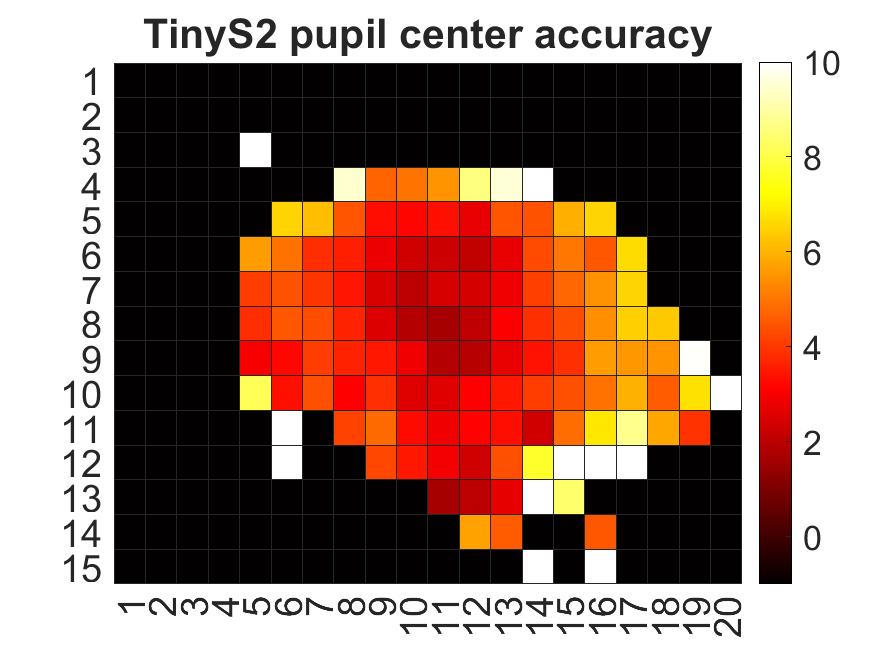}
	\includegraphics[width=0.24\textwidth]{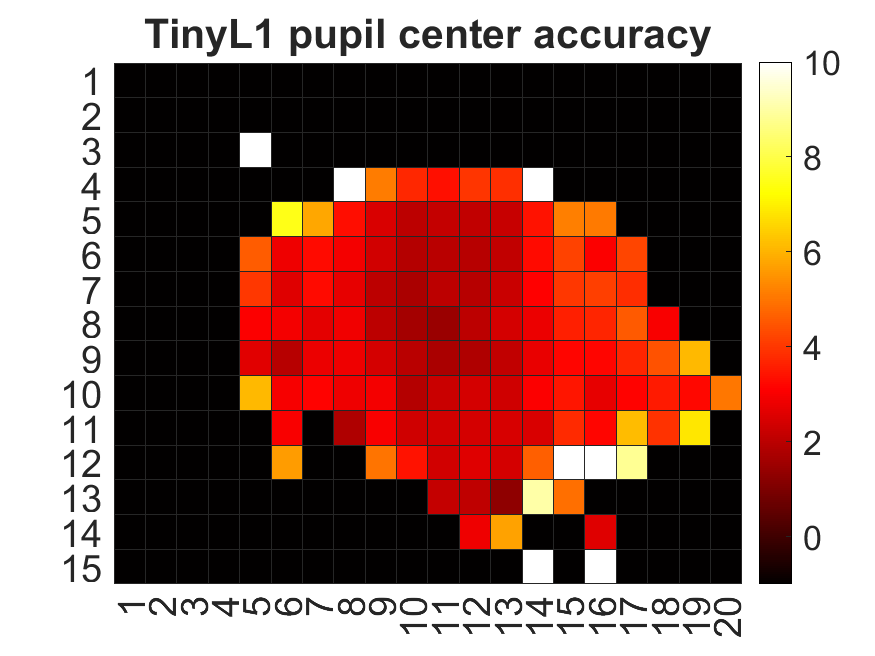}
	\includegraphics[width=0.24\textwidth]{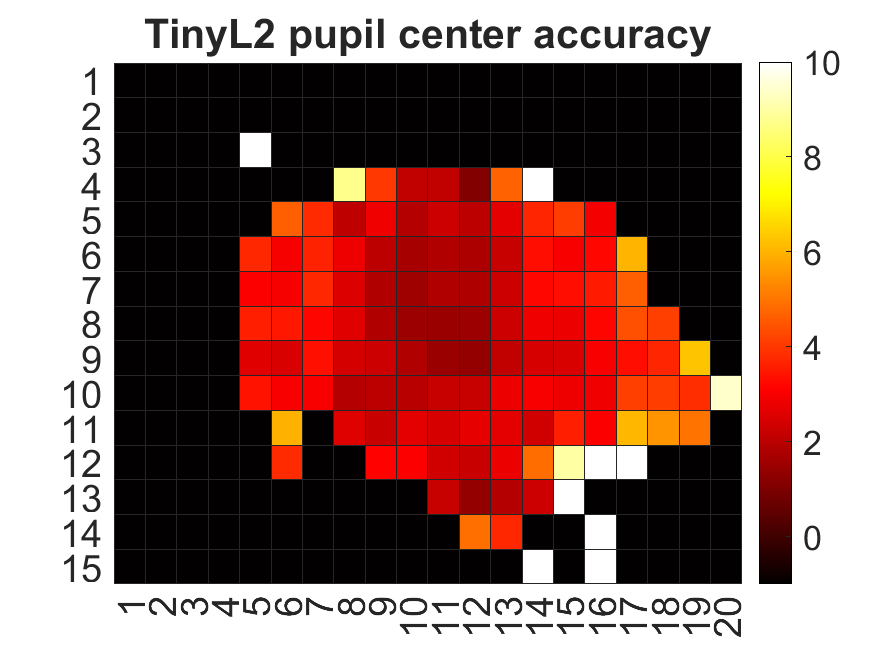}
	\includegraphics[width=0.24\textwidth]{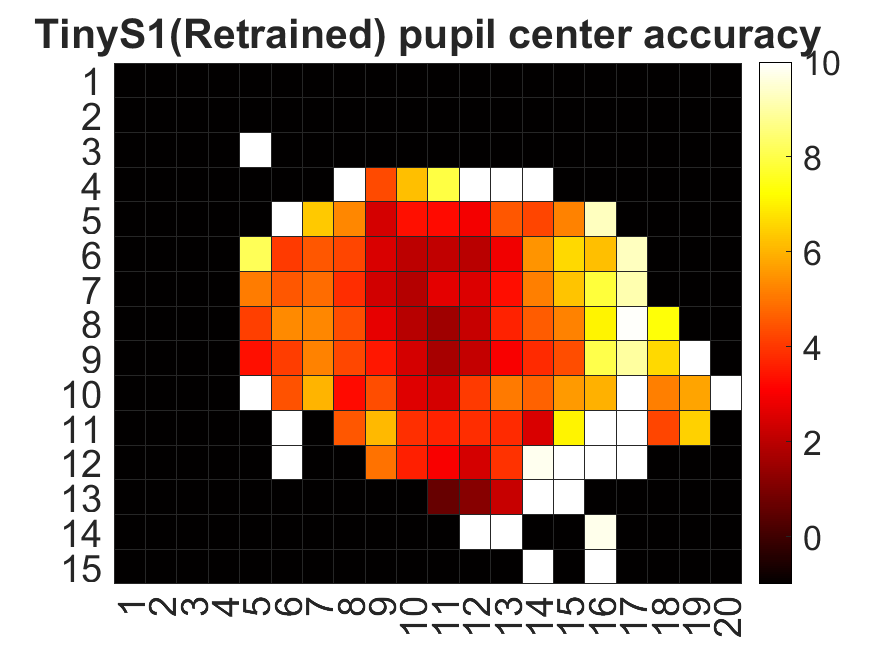}
	\includegraphics[width=0.24\textwidth]{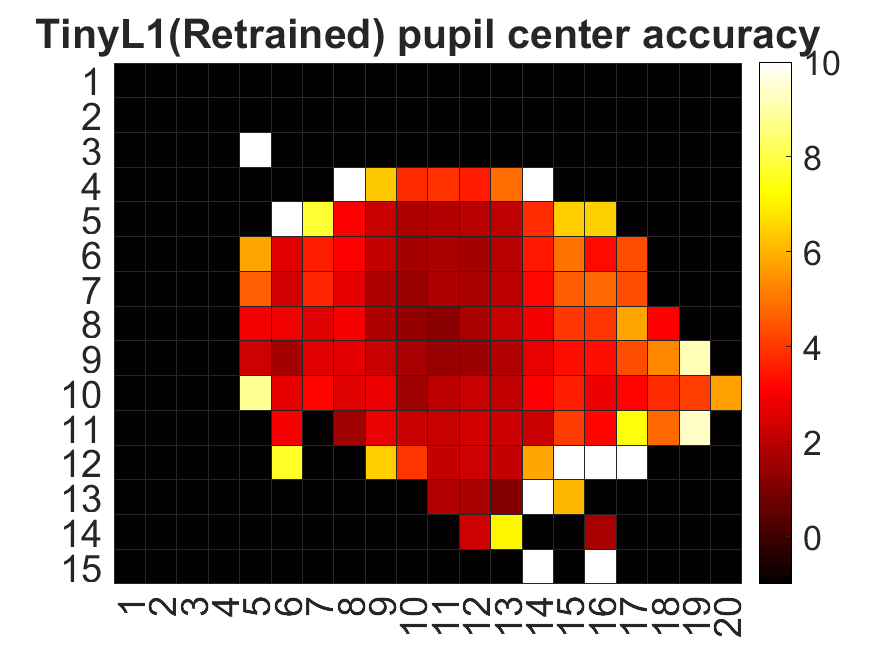}
	\includegraphics[width=0.24\textwidth]{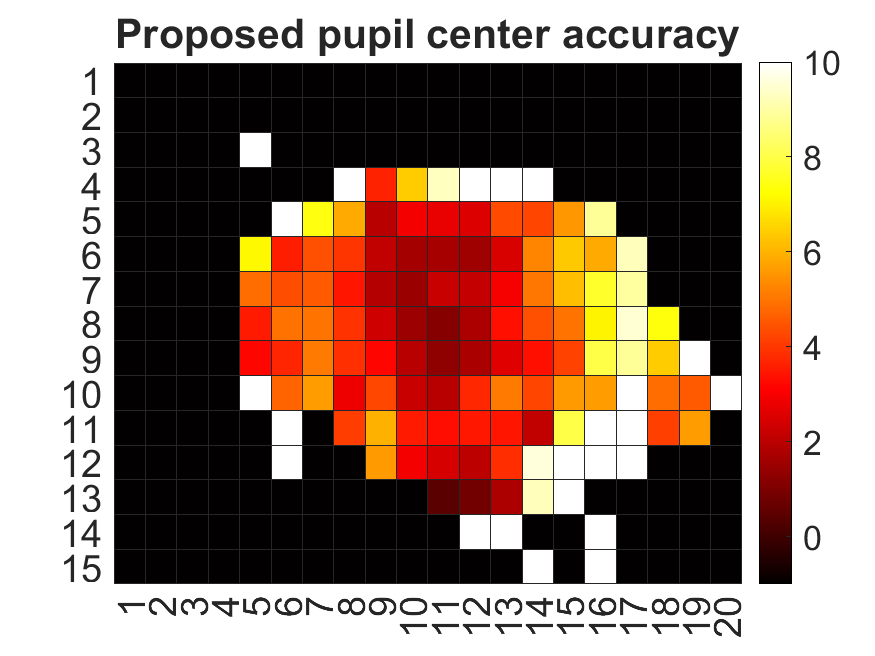}
	\includegraphics[width=0.24\textwidth]{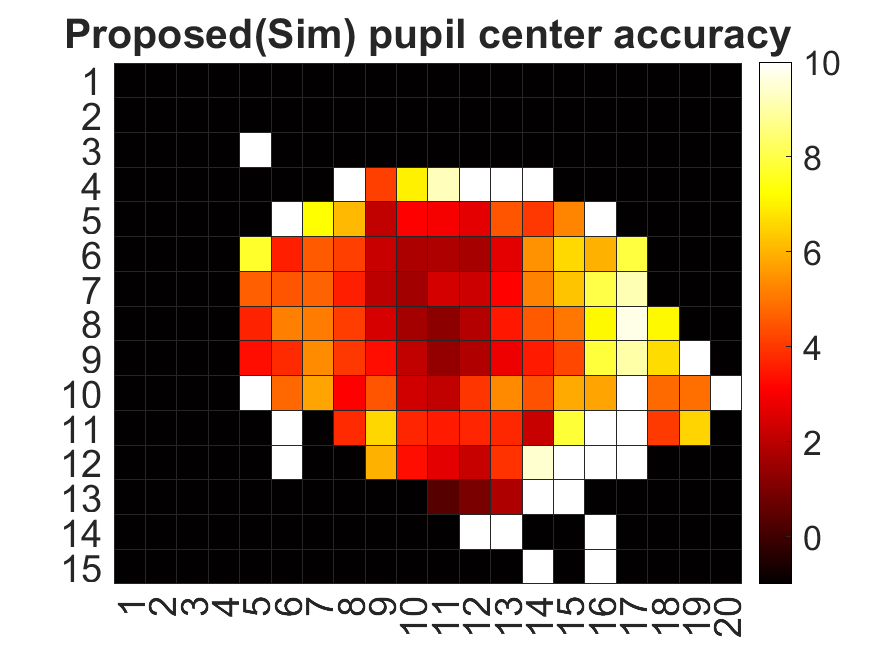}
	\caption{The mean euclidean pixel distance mapped to $10 \times 10$ cells on the image space. Higher values or brighter colors are a worse results. We have limited the maximum error to 10 pixels for a uniform display of the color scaling.}
	\label{fig:pcaoi}
\end{figure}

Figure~\ref{fig:cumuplot} on the left shows the cumulative mean intersection over union. This metric holds the information about the segmentation quality. On the right side of Figure~\ref{fig:cumuplot} is the cumulative euclidean pupil center pixel error which is important for the gaze estimation accuracy. As can be seen on both plots in Figure~\ref{fig:cumuplot}, SDM and BORE perform worse. BORE cannot handle the reflections very well as can be seen especially in Figure~\ref{fig:pcaoi} where a high mean pupil center error is present nearly everywhere on the image space. For SDM this is different since the method perform well in the near area of the mean shape (Figure~\ref{fig:pcaoi}). The best performance regarding a cumulative pupil center error of zero has ElSe (Figure~\ref{fig:cumuplot} right). It is also reaches the highest values for the cumulative intersection over union for a value of 0.9 (Figure~\ref{fig:cumuplot} left). Apart from this the TinyCNNs and the proposed approach are more robust and reaching nearly 90\% at a pixel error of two (Figure~\ref{fig:cumuplot} right). After the pixel error of two, our approch is outperformed by the TinyCNNs but our approach needs only a fraction of computation time (See Table~\ref{tab:stats}). For the segmentation quality, our approach keeps up with the TinyCNNs whereas the newly trained ones perform significantly better (Figure~\ref{fig:cumuplot} left). This is due to the reduction of possible ellipses as described in the method section. In addition, it can be seen in Figure~\ref{fig:cumuplot} that our approach trained on the simulated data performs only slightly worse compared to the one trained on the real data.

\begin{figure}
	\centering
	\includegraphics[width=0.24\textwidth]{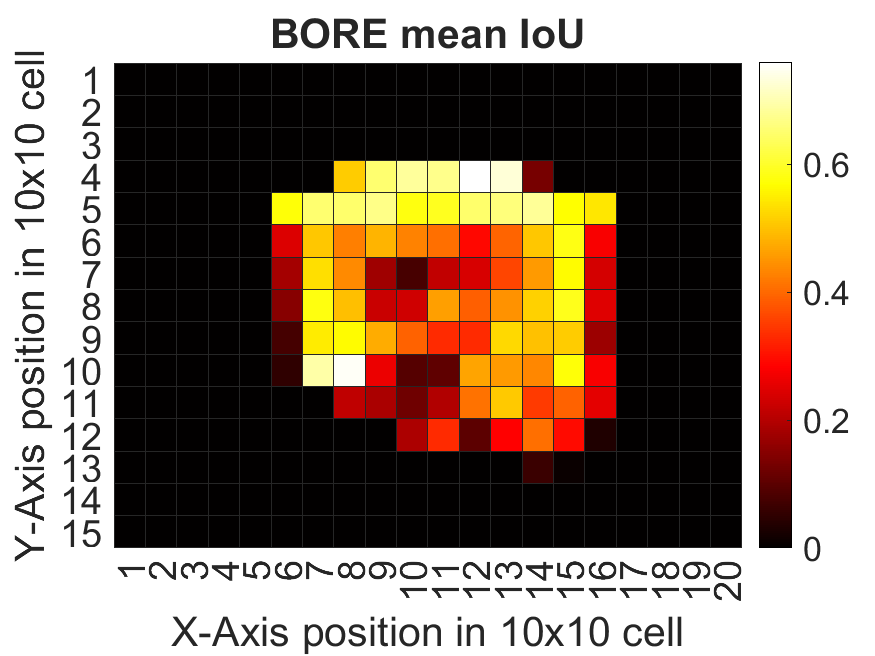}
	\includegraphics[width=0.24\textwidth]{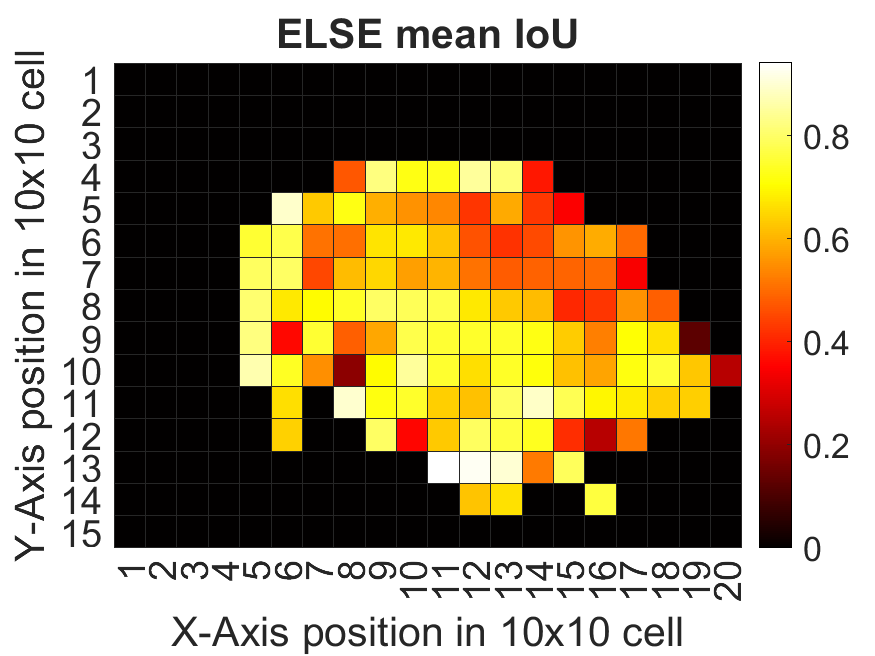}
	\includegraphics[width=0.24\textwidth]{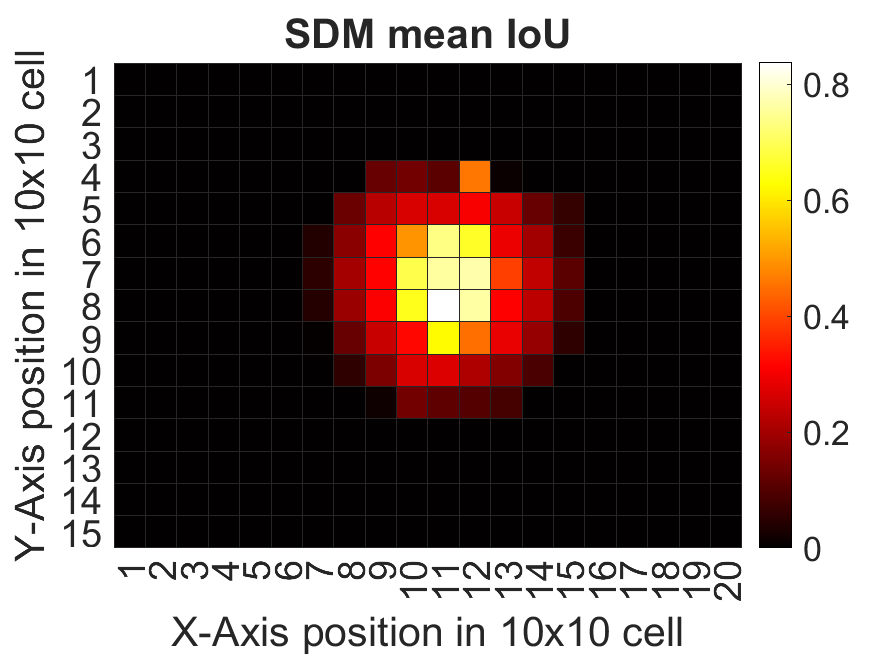}
	\includegraphics[width=0.24\textwidth]{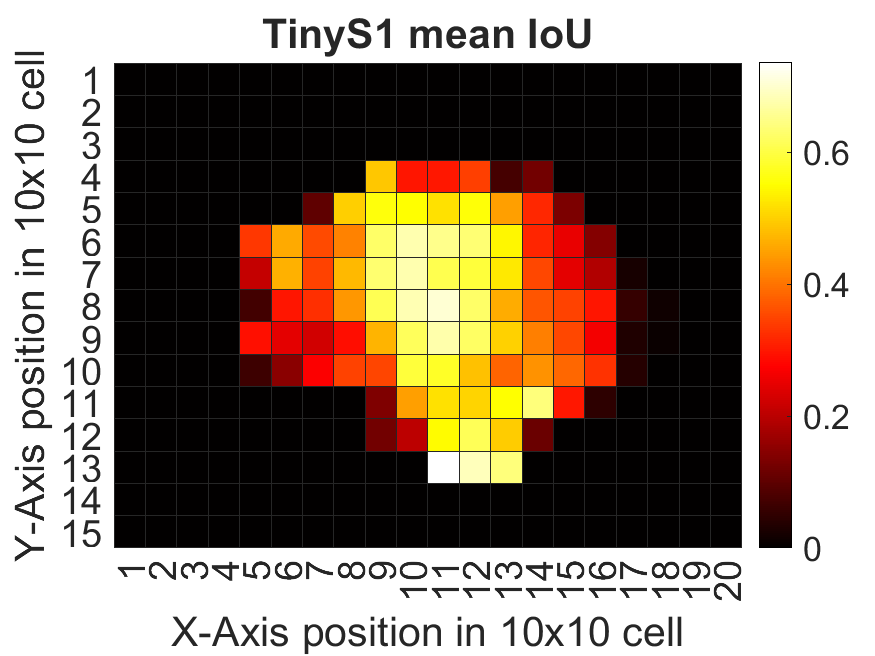}
	\includegraphics[width=0.24\textwidth]{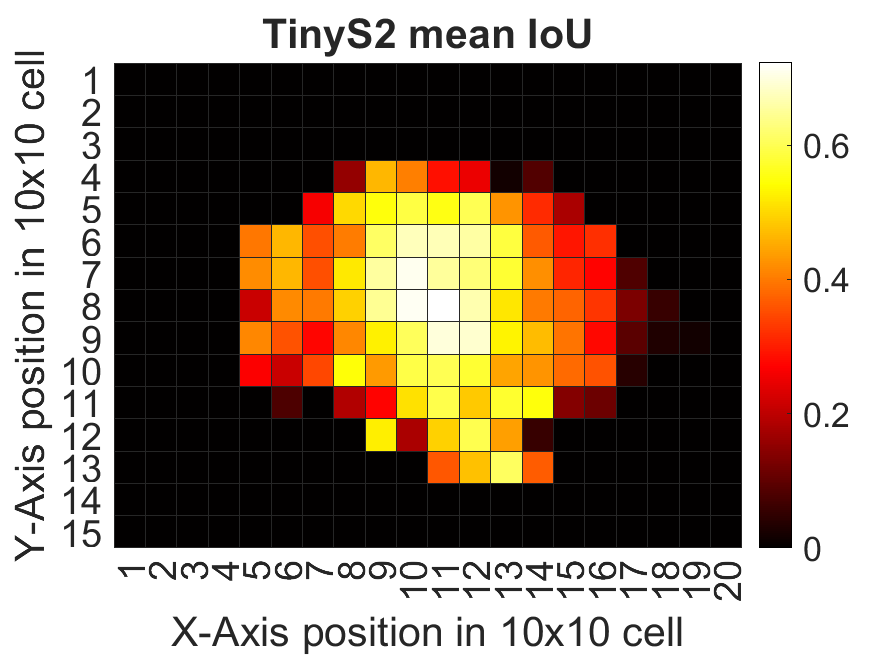}
	\includegraphics[width=0.24\textwidth]{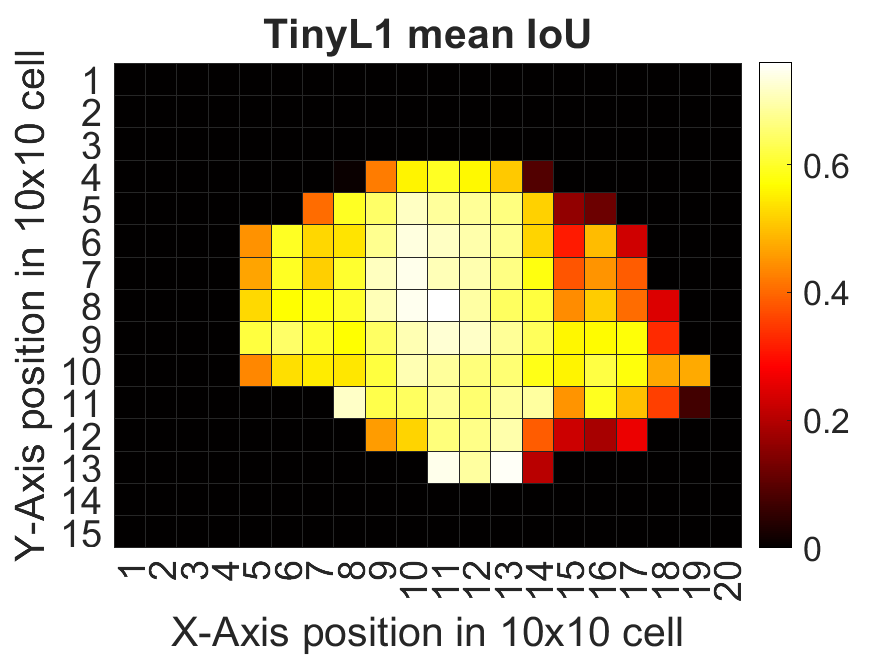}
	\includegraphics[width=0.24\textwidth]{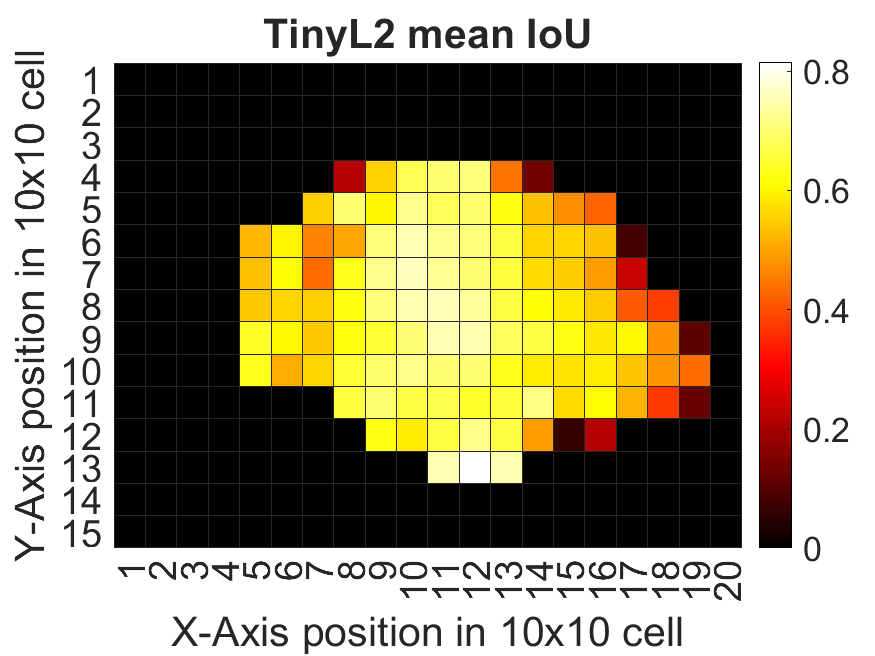}
	\includegraphics[width=0.24\textwidth]{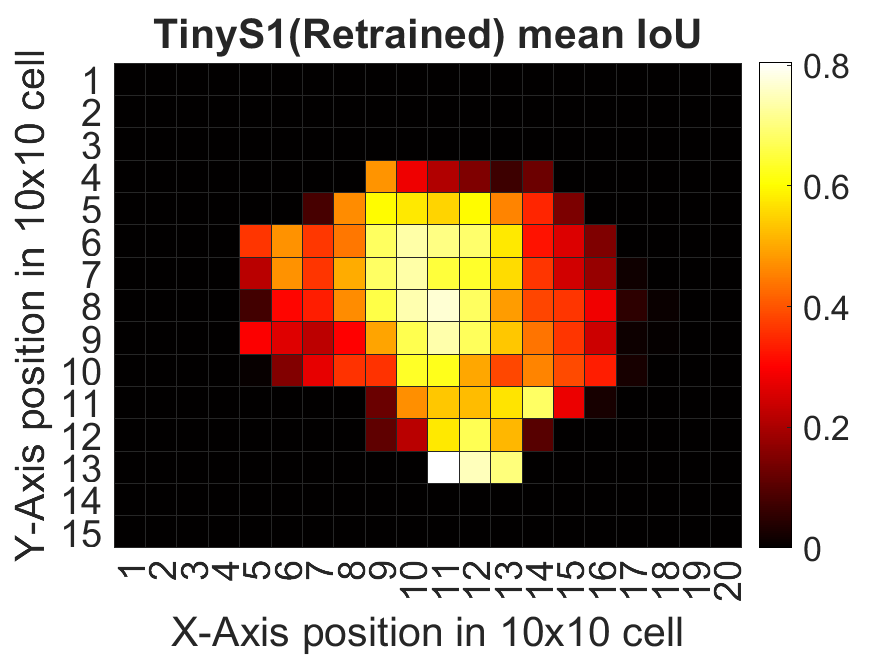}
	\includegraphics[width=0.24\textwidth]{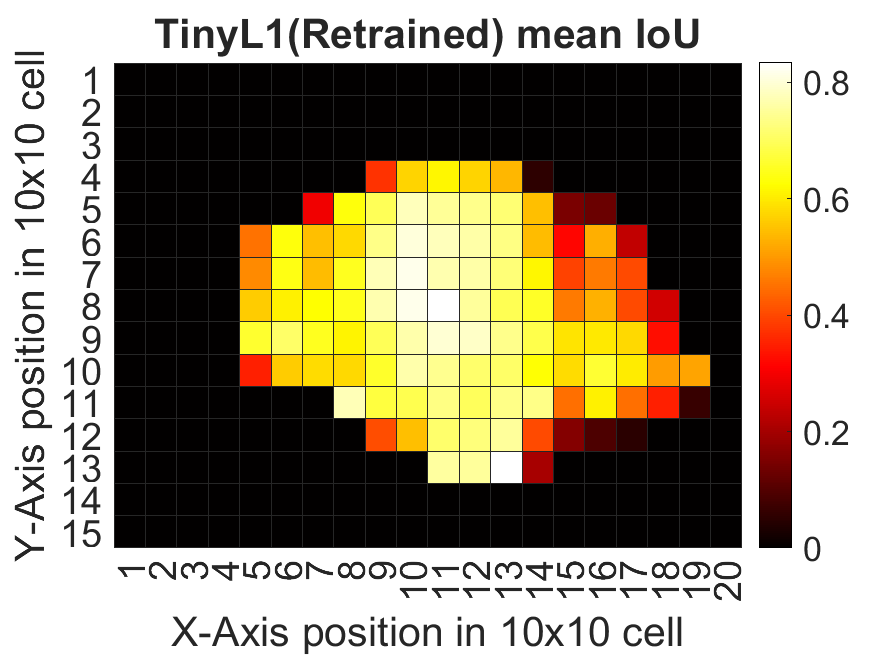}
	\includegraphics[width=0.24\textwidth]{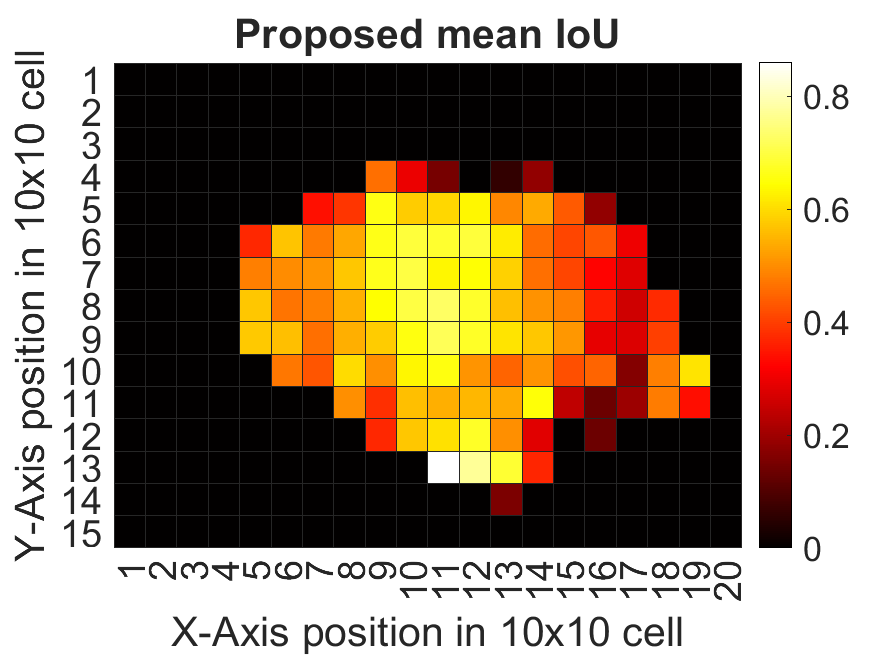}
	\includegraphics[width=0.24\textwidth]{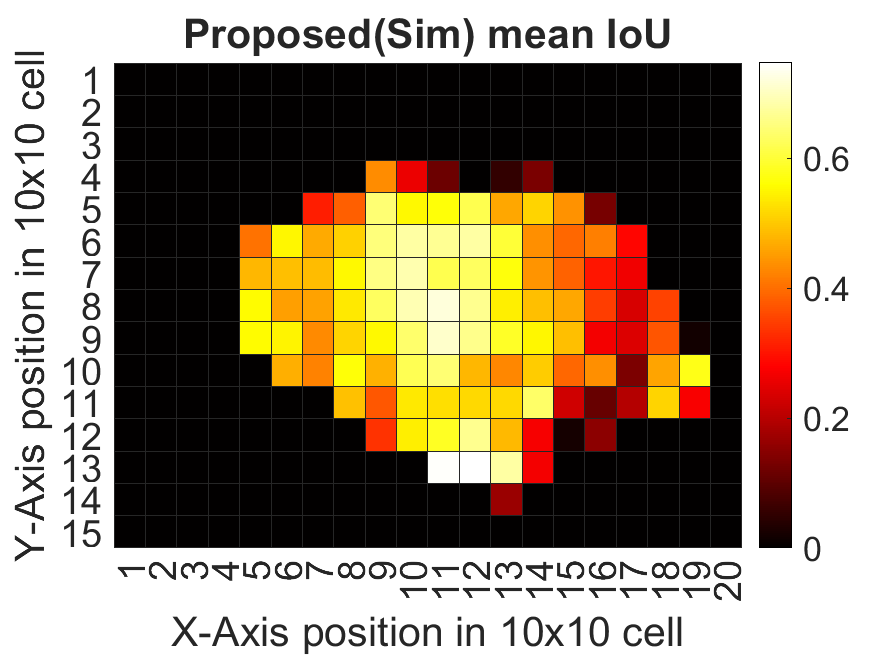}
	\caption{The mean intersection over union mapped to $10 \times 10$ cells on the image space. Higher values or brighter colors are better results. Please note that although the maximum is 1, the color scaling for all plots is not uniform with respect to the maximum. The minimum, on the other hand, is uniform at zero for all plots.}
	\label{fig:iouaoi}
\end{figure}

If Figure~\ref{fig:pcaoi} and Figure~\ref{fig:iouaoi} are compared for each approach, it can be seen that ElSe has a lot of invalid detections over the entire image space (Figure~\ref{fig:pcaoi}). This steams from heavy reflections which make edge detection not applicable. In Figure~\ref{fig:iouaoi} on the other hand, ElSe has a good average segmentation quality over the image space with the exception of the upper right area where occlusions by the eyelid occurred. Another important information is the clear center bias which can be seen for SDM by comparing Figure~\ref{fig:pcaoi} and Figure~\ref{fig:iouaoi}. Looking now at our approach and the TinyCNNs, we notice that they have good coverage of the entire image space (Figure~\ref{fig:pcaoi} and Figure~\ref{fig:iouaoi}). In terms of segmentation, however, our approach is significantly worse in the outer areas (Figure~\ref{fig:iouaoi}).

\begin{table}[h]
	\caption{Runtime for training and execution of the evaluated approaches on more than 2.500.000 images (5 augmented versions of each image) with data augmentation. The execution time was always evaluated with a single CPU core.}
	\label{tab:stats}
	\begin{tabular}{cccl}
		\toprule
		Method & Training time (h) & Execution (ms) & Note\\
		\midrule
		TinyCNN S1 & 40 (With GPU) & 3.7 & With teacher network training.\\
		TinyCNN S2 & 40 (With GPU) & 3.7 & With teacher network training.\\
		TinyCNN L1 & 44 (With GPU) & 7.9 & With teacher network training.\\
		TinyCNN L2 & 42 (With GPU) & 5.8 & With teacher network training.\\
		SDM (HOG+SVM) & 17 & 4.2 & \\
		BORE & 8 & 1.1 & Grid search for optimal pupil sizes.\\
		Else & 0 & 6.6 & No training necessary.\\
		Proposed & 1 & 0.9 & \\
		\bottomrule
	\end{tabular}
\end{table}

Table~\ref{tab:stats} shows the training time in hours as well as the execution time in milliseconds on a single CPU core. As can be seen, our approach outperforms the other approaches in terms of execution time. For the training time ElSe is the fastest approach due to it has not to be trained. Combining the detection, segmentation, training time, and execution time results we think the proposed approach is a valuable contribution to the eye tracking community.

\section{Limitations}
While the presented approach with Haar features in combination with statistical learning has a very low training time and a very low runtime and can also be trained on simulated data, this approach of course also has disadvantages. The first disadvantage is the search area. This means that no pupils or ellipses can be found outside this area. Of course, this limitation can be easily circumvented by arbitrarily extending the search area, but this has a negative impact on both the detection rate and the runtime. Another disadvantage of the presented approach, is the statistic itself, which in the case of feature weighting weights frequent occurrences of valid features more heavily. This means that large data sets of similar images lead to features that are valid in these images being weighted more heavily than others. This results in an overfitting to these images. Also, the presented approach only recognizes shapes which are also present in the training data. This is because unknown shapes are not sampled and have no probability of occurrence. This can be easily fixed by simulated data or data manipulation, but this also leads to an increased runtime.

\textit{\textbf{How we think the algorithm should be applied:} Since the presented algorithm can be used very performantly and statistical learning can be used very efficiently for training, our idea for the application is a direct training after calibration. Here, an expensive deep neural network could be used in the first step to segment the pupils offline. Then statistical learning is used to weight the haar features and the ellipses. Through this, it would be possible to create a personalized detector, as is the case with BORE~\cite{ETRA2018FuhlW}, and deploy it online in a resource-efficient manner. A disadvantage of this approach is, of course, that the one-point calibration could not be used in this case but a coverage of the whole area would have to be guaranteed.}

\section{Conclusion}
In this work, we have presented a new approach to efficiently train and segment pupils. While it is not able to segment pupils as accurately as, for example, edge-based approaches, it is comparatively robust and very efficient to compute. To overcome the disadvantage of segmentation quality, finer segmentation can of course be performed in a second step but this again incurs additional computational overhead. Overall, we believe that our approach is a valuable contribution to the online adaptation of pupil detectors and their use in high speed eye tracking.

\bibliographystyle{ACM-Reference-Format}
\bibliography{sample-base}

\end{document}